\RequirePackage{fix-cm}
\documentclass[twocolumn,epjc3]{svjour3}  
\smartqed  
\usepackage{lipsum}
\usepackage{amssymb,amsmath}
\usepackage{comment}
\usepackage{graphicx}
\usepackage{nccmath}
\usepackage{float}
\usepackage{mathrsfs}
\journalname{Eur. Phys. J. C}
\usepackage[backend=biber,style=ieee,citestyle=numeric-comp,sorting=none]{biblatex}
\usepackage{etoolbox}

\AtBeginBibliography{\sloppy}
\begin{document}

\title{Testing conformal gravity using the supermassive black hole NGC 4258
}
\subtitle{}


\author{Diego A. Martínez-Valera\thanksref{e1,addr1}
        \and
        Alfredo Herrera-Aguilar\thanksref{e2,addr1} 
}

\thankstext{e1}{e-mail: dmartinez@ifuap.buap.mx}
\thankstext{e2}{e-mail: aherrera@ifuap.buap.mx}


\institute{Instituto de Física, Benemérita Universidad Autónoma de Puebla, Apartado Postal J-48, 72570 Puebla, Puebla, México \label{addr1}
}

\date{Received: date / Accepted: date}

\maketitle

\begin{abstract}
In this paper, we perform a Bayesian statistical fit to estimate the free parameters of a nonsingular black hole in conformal gravity by employing megamaser astrophysical data of the supermassive black hole hosted at the center of the active galactic nucleus of NGC 4258. This estimation has been carried out by taking into account a general relativistic approach, which makes use of the positions on the sky of the photon sources and the frequency shift observations from the water megamaser system in circular motion around the black hole. Within the framework of conformal gravity, a way to eliminate the singularity at $r=0$ from the Schwarzschild spacetime is by introducing a conformal factor characterized by a length scale parameter $l$ and an integer parameter $N$. Therefore, the spacetime geometry depends on the mass of the black hole, and the conformal gravity parameters $l$ and $N$. In this work, we estimate the mass-to-distance ratio $M/D$ and the length scale ratio $l/D$ with fixed values of the integer parameter $N=1,2$, considering the geodesics of conformally/non-conformally coupled massive particles. This method leads to posterior Gaussian distributions for all  parameters, thus yielding a most probable value for the parameter $l$ for both conformally/non-conformally coupled particles, in contrast to previous constraints based on X-ray astrophysical data, where an upper bound for the parameter $l$ has been established. Furthermore, we obtain new physical properties regarding the existence of the $ISCO$ radius for this nonsingular spacetime in the case of non-conformally coupled particles.
\end{abstract}

\section{Introduction}
\label{introduction}

Einstein's theory of gravity is the most accurate theory for describing gravitational phenomena so far, and has been successful over many observational tests since its proposal in 1915. One of the main predictions of the theory is the existence of black holes (BHs), which have been extensively studied since the second half of the 20$^\text{th}$ century, and during the last decades we have witnessed an unprecedented advance in the understanding of BH physics. The existence of such enormously dense objects in nature has been recently supported by observations of the shadow images of supermassive black holes (SMBHs) hosted at the core of M87 and Milky Way galaxies accomplished by the Event Horizon Telescope Collaboration \cite{EHT1,EHT2,EHT3}, by gravitational wave detections caused by the merging of two stellar-mass BHs performed by the LIGO-Virgo collaborations \cite{LIGO1,LIGO2}, by  observations of star motion around the center of the Milky Way \cite{Ghez1,Ghez2,Genzel1,Genzel2}, as well as by frequency shift measurements of the accretion disk maser-emission lines of water molecules orbiting SMBHs in active galactic nuclei (AGNs) \cite{Moran, Lo}, to mention a few examples.

These pieces of evidence have served as a motivation to develop methods for testing gravity theories and obtaining information on the BH background spacetime, thus acquiring a deeper understanding of the underlying physics. Among all these efforts, a remarkable one is the general relativistic method developed in \cite{HN-1,HN-2,KdS}, which allows one to obtain the BH parameters in terms of the observational total frequency shift of light emission from probe particles orbiting a BH. Besides, in comparison to the Newtonian approach, the general relativistic method allows one to account for the gravitational redshift due to the curvature of spacetime apart from the kinematic frequency shift generated by the Doppler effect, \textit{i.e.} by the probe particles' motion. This method provides a more accurate description of the real phenomena when using data coming from test particles orbiting near enough the BH, so that general relativistic effects can be taken into account. Hence, the ideal laboratory for testing strong gravity is the spacetime around astrophysical BHs. The aforementioned method has provided a way to estimate the mass-to-distance ratio for 16 SMBHs located at the core of AGNs with water vapour clouds in its accretion disk \cite{ApJLNucamendi,Villalobos1,Villaraos,Villalobos2,AGJ}. These estimations have been performed by using the frequency shift from  water megamasers, which consist of vapour clouds emitting at 22 GHz by stimulated emission with a considerable luminosity ($L>100\,L_{\odot}$). The redshift water masers emission was first detected in several AGNs \cite{Claussen,Nakai}. In particular, the NGC 4258 megamaser system has been thoroughly studied, and it is considered the archetypal model for megamaser configurations \cite{Moran, Lo}, with almost symmetric bands of features about its systemic velocity of circa $500$ km s$^{-1}$, with highly redshifted/blueshifted features with velocities near $1500$/$-500$ km s$^{-1}$. Furthermore, high precision Very Long Baseline Interferometry (VLBI) information provides evidence for the source of this spectral structure to be a rapidly rotating disk modeled by Keplerian motion, implying early estimations for the central mass of the order of $10^{7}M_{\odot}$. Further estimations rendered a value for the mass of around $4 \times 10^{7} M_{\odot}$ \cite{DistanceNGC4258} and a geometric distance from the Earth to NGC 4258 of $D=7.576\pm 0.082\,(\text{stat.}) \pm 
 0.076\,(\text{sys.})$ Mpc \cite{DistanceNGC4258,ImprovedNGC4258}. Very Long Baseline Array (VLBA) observations showed a distribution of the masers extending from 0.17 to 0.28 pc on either side of the systemic masers \cite{NGC4258}.

On the other hand, BHs in extended theories of gravity have been extensively investigated in recent times. Taking into account gravitational wave and electromagnetic detections from BHs, we find that there are still uncertainties in estimating their parameters \cite{EHT1,EHT2,EHT3,LIGO1,LIGO2,Ghez1,Ghez2,Genzel1,Genzel2}. This fact allows for the existence of BH solutions beyond Einstein theory of relativity and leads to the opportunity for exploring their strong field regime. These solutions contain new free parameters characterizing the gravitational model and consequently, further information on the underlying extended theory of gravity can be obtained from analyzing the frequency shift of photons emitted by astrophysical sources and detected observations here on the Earth.

In this direction, we have applied the method proposed in \cite{HN-1,HN-2} to a metric solution belonging to an extended theory of gravity, namely, to conformal gravity, and we have obtained the total frequency shift of photons for the case of a nonsingular BH. In addition to the mass-to-distance ratio $M/D$, this BH is characterized by the ratio $l/D$, where $l$ is a parameter with dimension of length, and an integer parameter $N$ \cite{MartínezV}. Henceforth, in this paper we perform an estimation of these parameters using the observational data from the megamaser system NGC 4258, to constrain the values for $l/D$ and $M/D$, by considering the cases where $N=1$ and $N=2$, for simplicity.

We have chosen the NGC 4258 spiral galaxy to perform this case study because it provides one of the strongest and more compelling astrophysical evidences for the existence of SMBHs at the core of AGNs due to its highly accurate measurement of gas dynamics, constituting one of the most robust megamaser systems discovered so far. Moreover, it is statistically very appealing since it possesses a large data set of measurements of highly frequency shifted maser-emission lines of water vapour on the accretion disk, with nearly 14,000 maser features, among which 10,039 are systemic, 3979 redshifted, and 289 blueshifted masers. It is rather difficult to find maser systems, with only about a hundred systems found so far, possibly due to the fact that only megamaser systems with accretion disks viewed edge-on from Earth can be detected. The NGC 4258 system is particularly good because of its dynamical properties, namely, it provides the first direct evidence of an accretion disk in which the gas orbits the central object in nearly circular paths. The presence of $\text{H}_{2}\text{O}$ megamasers (which can be mapped at sub-milli-arc-second resolution) represents a powerful tool to probe spatial and kinematic distributions of molecular gas in distant galactic nuclei at scales below 1 pc \cite{Moran,Lo}. Additionally, it is one of the few megamaser systems that are not within the Hubble flow, which allows us to apply the formalism developed in \cite{MartínezV}, where the accelerated expansion of the universe is not taken into account. Nevertheless, the peculiar motion of the galaxy NGC 4258 with respect to our galaxy must be considered, and it can be described by a special relativistic boost $1+z_{boost}$ (see Section \ref{GR model}).

This study presents a novel treatment of the statistical estimation of the parameter $l$, in comparison to previous fits, where conformal gravity has been tested by using the iron K$\alpha$ method of an accretion disk in a BH spacetime \cite{TestCG1} resulting in an upper bound for the ratio $l/M<1.2$, while a more stringent constraint $l/M<0.45$ was obtained by employing the X-ray reflection spectroscopy method using astrophysical data from the SMBH hosted at the center of the galaxy 1H0707-495 \cite{TestCG2}. The most recent constraint $l/M<0.12$ was performed by using astrophysical data from the stellar-mass BH hosted at the center of  the GS 1354–645 system \cite{TestCG3}. It is worth mentioning that the constraints found in these articles are for the parameter $l/M$, whereas in our work we have estimated the parameter $l/D$ due to the coupling that exists between the length scale parameter $l$ and the radius of the emitter $r_e$ in the equations for the frequency shift. We also estimate the ratio $M/D$, a fact that allows us to estimate the ratio $l/M$ upon the corresponding error propagation.

This paper is outlined as follows. In the next Section we present a brief review of the general relativistic method applied to nonsingular BH in conformal gravity considering two cases, namely, conformally and non-conformally coupled massive probe particles. In Section \ref{Observations} we briefly review the observations of the NGC 4258 megamaser system reported from 3 years 18 epoch VLBI observations. In Section \ref{nonsingular} we present the statistical fit for the megamaser data of NGC 4258 using the models presented in Section \ref{GR model} and we discuss the main results. In the last Section, we present some final remarks.

\section{General relativistic model for a nonsingular BH in conformal gravity}
\label{GR model}
Here we present a review of the general relativistic model from which we obtain the total frequency shift of light emitted by probe particles circularly orbiting a nonsingular BH in conformal gravity \cite{Modesto16,Bambi17}. In order to provide a thorough study of the frequency shift of BHs within conformal gravity, we examine two cases of probe particles: a non-conformally coupled particle, for which we assume that conformal symmetry has already been broken at the free particle's lagrangian level; and a conformally coupled probe particle, where we explicitly break conformal symmetry by fixing the gauge in the final expression of the total frequency shift. The difference between these two approaches lies in the stage at which we break the conformal symmetry during the derivation of the geodesics.

The spacetime of such nonsingular BH is described by the following metric 
\begin{equation}
    \begin{split}
ds^2&=g_{\mu\nu}dx^{\mu}dx^{\nu}=\Omega(r)\hat{g}_{\mu\nu}dx^{\mu}dx^{\nu}=\left(1+\frac{l^2}{r^2}\right)^{2N}\\
&\times\left[\left(1-\frac{2M}{r}\right)dt^2+\left(1-\frac{2M}{r}\right)^{-1}dr^2+r^2d\xi^2\right],
\end{split}\label{CG metric}
\end{equation}
where $g_{\mu\nu}$ is the conformally rescaled metric, $\hat{g}_{\mu\nu}$ stands for the Schwarzschild metric, $\Omega(r)=\left(1+\frac{l^2}{r^2}\right)^{2N}$ is the conformal factor characterized by the parameters $l$ and $N$, and $d\xi^2=d\theta^2+\sin^2\theta d\varphi^2$. Here, we have performed a conformal transformation to the Schwarzschild spacetime, thus removing the singularity at $r=0$.

This metric comes from a super-renormalizable theory which is made conformally invariant by introducing a compensator scalar field that multiplies the \textit{background metric} and it is closely related to the conformal factor \cite{Modesto16}. Therefore, the theory can be analyzed from two different frames, and the background metric (in this case, the Schwarzschild spacetime) multiplied by the conformal factor, gives rise to a \textit{physical metric} (see the Appendix for further details). Thus, from the point of view of the physical metric, the scalar field sector in the non-physical frame encodes the geometrical modifications allowing for a nonsingular spacetime. Therefore, the spacetime metric given by \eqref{CG metric} represents a solution for both a weakly non-local (or quasi-local) quadratic theory in the Riemann curvature and the finite conformal gravity.

Additionally, it is worth mentioning that when we consider a rotating black hole, an expansion of the total frequency shift expressions for large orbits of the emitter, and nontrivial peculiar redshift, reveals that the angular momentum appears up to the eleventh term, and therefore it becomes completely negligible (see Eqs. (38) and (39) in \cite{HN-2} and discussion therein). In this sense, using a spherically symmetric spacetime to study the frequency shift of photon emission from megamaser systems is definitely suitable.

\subsection{Non-conformally coupled probe particles}
 
For the study of the geodesic motion of a non-conformally coupled (NCC) massive test particle, we assume that conformal invariance is broken. Thus, the particles' motion carries no additional information of the former symmetry, where the only remnant of it is encoded in the conformal factor. Therefore, our analysis begins with the geodesic equation
\begin{eqnarray}
    g_{\mu\nu}U^{\mu}U^{\nu} = -1,
\end{eqnarray}
where $U^{\mu}$ is the 4-velocity of the probe particle.

Furthermore, the metric \eqref{CG metric} represents a static and spherically symmetric background that possesses two commuting Killing vectors associated with the conserved quantities along the circular geodesic motion of the probe particles. These conserved quantities are the energy per unit mass $E$ and the axial angular momentum per unit mass $L$. In addition, due to the spherical symmetry of the line element \eqref{CG metric}, we restrict the particle's motion to the equatorial plane $\theta =\pi /2$ without loss of generality. Besides, by considering the particle's motion with normalized 4-velocity $U^{\mu}U_{\mu}=-1$, constrained to equatorial motion where $U^\theta =0$, we obtain an equation with an energy conservation law structure
\begin{equation}
    -\frac{1}{2}g_{tt}g_{rr}\left(U^{r}\right)-\frac{1}{2}g_{tt}\left( 1+\frac{L^2}{g_{\varphi\varphi}}\right) = \frac{E^2}{2}, \label{Conservation law}
\end{equation}
where the first term is the kinetic energy of the particle moving in an effective potential of the form
\begin{equation}
V_{eff}=-\frac{1}{2}g_{tt}\left( 1+\frac{L^2}{g_{\varphi\varphi}}\right).
\label{Eff Pot}
\end{equation}

When taking into account real astrophysical systems with spherical symmetry in the spacetime background, investigating circular equatorial motion is relevant. Hence, the effective potential $V_{eff}$ must satisfy the following conditions for circular orbits
\begin{equation}
V_{eff}=\frac{E^2}{2}, \,\,\,\,\, V'_{eff}=0,
\end{equation}%
where the prime symbol denotes the derivative with respect to the radial coordinate $r$. For stability of the circular orbits, it is required for the second derivative of the effective potential to satisfy the following condition 
\begin{eqnarray}
V''_{eff}\geq 0,  \label{Stability}
\end{eqnarray}%
where the equality yields the innermost stable circular orbit (ISCO).

On the other hand, the expression for the frequency shift $\omega$ in static and spherically symmetric backgrounds of the form (\ref{CG metric}) for circular equatorial orbits ($U^{r}=0$ and $U^{\theta }=0$), can be written as \cite{HN-1,HN-2,MartínezV}
\begin{equation}
1+z_{_{1,2}}\!=\frac{\omega _{e}}{\omega _{d}}=\frac{\left(k_{\mu}U^{\mu}\right)|_e}{\left(k_{\mu}U^{\mu}\right)|_d}
= \frac{\left(U_{e}^{t}-b_{\gamma}U_{e}^{\varphi }\right)|_e}{\left(U_{e}^{t}-b_{\gamma}U_{e}^{\varphi }\right)|_d },   \label{GeneralShift}
\end{equation}
where $b_{\gamma}$ is the light bending parameter defined as $b_{\gamma}=L_{\gamma}/E_{\gamma}$, where $E_{\gamma}$ and $L_{\gamma}$ are the energy and angular momentum of the emitted photons, respectively. The $b_{\gamma}$ parameter is a conserved quantity along geodesics that maximizes at the points $\varphi\approx\pm \pi/2$, where $k^{r}=0$, and has the following explicit form for the spacetime \eqref{CG metric}
\begin{eqnarray}
    b_{\gamma _{\mp }}&= \mp \sqrt{\frac{r^3}{r-2M}}. \label{deflectPar}
\end{eqnarray}

In addition, the nonvanishing 4-velocity components of the NCC massive probe particle in circular and equatorial motion read
\begin{equation}
    U^t_e=\sqrt{\frac{r_e(r_e^2+l^2)-2Nl^2r_e}{\Omega(r_e-3M)(r_e^2+l^2)}},
    \label{U^t}
\end{equation}
\begin{equation}
    U^{\varphi}_e=\pm\sqrt{\frac{M(r_e^2+l^2)-2Nl^2(r_e-2M)}{\Omega(r_e-3M)(r_e^2+l^2)r_e^2}},
    \label{U^phi}
\end{equation}
where the $\pm$ sign indicates clockwise or counterclockwise direction of motion with respect to a far away observer. 
Furthermore, by considering the observer at a very large distance from the BH, its 4-velocity is given by $U_{d}^{\mu }=(1,0,0,0)$, and therefore the expression for the frequency shift \eqref{GeneralShift} reduces to 
\begin{equation}
1+z_{_{1,2}}\!=U_{e}^{t}-b_{\gamma _{\mp }}U_{e}^{\varphi }.\label{shift2}
\end{equation}

 Hence, from Eqs. \eqref{deflectPar}, \eqref{U^t}, and \eqref{U^phi}, the total frequency shift in the conformal BH background reads%
\begin{equation}
\begin{split}
    1+z_{_{1,2}}^{NCC}&=\sqrt{\frac{r_e}{\Omega(r_e-3M)}}\\
    &\times\left(\sqrt{1-\frac{2Nl^2}{r_e^2+l^2}} \pm \sqrt{\frac{M}{r_e-2M}-\frac{2Nl^2}{r_e^2+l^2}}\right), \label{Shift1}
\end{split}
\end{equation}
where one identifies the gravitational frequency shift
\begin{equation}
\begin{split}
    1+z_{g}^{NCC}&=\sqrt{\frac{r_e}{\Omega(r_e-3M)}\left(1-\frac{2Nl^2}{r_e^2+l^2}\right)},\label{zgrav}
\end{split}
\end{equation}
and the kinematic frequency shift
\begin{equation}
\begin{split}
    z_{kin_{\pm }}^{NCC}&=\pm\sqrt{\frac{r_e}{\Omega(r_e-3M)}\left(\frac{M}{r_e-2M}-\frac{2Nl^2}{r_e^2+l^2}\right)},
\end{split} \label{zkin}
\end{equation}
satisfying the relation 
\begin{equation}
z_{_{1,2}}=z_{g}+z_{kin_{\pm}}.
\label{ztotal}
\end{equation}

\subsection{Conformally coupled particles}

We now begin the analysis of conformally coupled (CC) massive particles with the lagrangian of a free particle that preserves conformal invariance \cite{Modesto16, Bambi17}
\begin{equation}
    \mathscr{L}= -\sqrt{-(f\phi^{*})^{2}g_{\mu\nu}\dot{x}^{\mu}\dot{x}^{\nu}},\label{Lagrangian}
\end{equation}
where $\phi^{*}$ is an auxiliar scalar field that helps to achieve conformal invariance, the overdot indicates derivative with respect to the proper time $\tau$, and $f$ is a positive constant. The corresponding Euler-Lagrange equations lead to the following relations
\begin{equation}
    \frac{\partial \mathscr{L}}{\partial \dot{t}} = -\frac{(f\phi^{*})^{2}g_{tt}\dot{t}}{\mathscr{L}} = const. = -E,
\end{equation}
and
\begin{equation}
    \frac{\partial \mathscr{L}}{\partial \dot{\varphi}} = -\frac{(f\phi^{*})^{2}g_{\varphi\varphi}\dot{\varphi}}{\mathscr{L}}  = const. = L,
\end{equation}
where $E$ and $L$ are the energy and the axial angular momentum, respectively. Furthermore, identifying that $U^{t} = \dot{t}$,  $U^{\varphi} = \dot{\varphi}$, and  $g_{\mu\nu}\dot{x}^{\mu}\dot{x}^{\nu} = -1$, we rewrite the former expressions as follows
\begin{equation}
    U^{t} = -\frac{E}{f\phi^{*}g_{tt}}, \label{t-4Velocity}
\end{equation}
\begin{equation}
    U^{\varphi} = \frac{L}{f\phi^{*}g_{\varphi\varphi}}. \label{phi-4Velocity}
\end{equation}

Thus, if we expand the geodesic equation and we substitute $U^{t}$ and $U^{\varphi}$ in terms of $E$ and $L$, while setting the motion to the equatorial plane $\theta =\pi/2$ ($U^{\theta} = 0$), we obtain the following expression
\begin{equation}
    \frac{E^2}{2} = -\frac{\phi^{*2}g_{tt}}{2}\left(1+\frac{L^2}{\phi^{*2}g_{\varphi\varphi}}\right) - \frac{\phi^{*2}g_{tt}g_{rr}}{2}\left(U^{r}\right)^{2},
\end{equation}
where we have redefined the energy and angular momentum as $E/f \rightarrow  E$ and $L/f\rightarrow L$. The equation above takes the form of an energy conservation law analogous to Eq. \eqref{Conservation law}, with an effective potential
\begin{equation}
    V_{eff}(r) = -\frac{\phi^{*2}g_{tt}}{2}\left(1+\frac{L^2}{\phi^{*2}g_{\varphi\varphi}}\right).
\end{equation}

Then, if we define $\tilde{g}_{tt} = \phi^{*2}g_{tt}$, and $\tilde{g}_{\varphi\varphi} = \phi^{*2}g_{\varphi\varphi}$,
 Eqs. \eqref{t-4Velocity} and \eqref{phi-4Velocity} become
\begin{equation}
    U^{t} = -\phi^{*}\frac{E}{\tilde{g}_{tt}}, \label{t-4Velocity2}
\end{equation}
\begin{equation}
    U^{\varphi} = \phi^{*}\frac{L}{\tilde{g}_{\varphi\varphi}}. \label{phi-4Velocity2}
\end{equation}
Thus, we can use the results of \cite{MartínezV} to obtain the non-zero 4-velocities in terms of the metric components as follows
\begin{equation}
    U^{t} = \phi^{*}\sqrt{\frac{\tilde{g}_{tt}^2}{\tilde{g}'_{tt}\tilde{g}_{\varphi\varphi}-\tilde{g}_{tt}\tilde{g}'_{\varphi\varphi}}},
\end{equation}
\begin{equation}
    U^{\varphi} = \phi^{*} \sqrt{\frac{\tilde{g}_{\varphi\varphi}^2}{\tilde{g}_{tt}\tilde{g}'_{\varphi\varphi}-\tilde{g}'_{tt}\tilde{g}_{\varphi\varphi}}},
\end{equation}
and by performing a similar analysis to the one presented in the previous Section, we obtain the total frequency shift
\begin{equation}
1+z_{_{BH_{1,2}}}=\frac{\phi^{*}}{\sqrt{\tilde{g}_{\varphi \varphi }\tilde{g}_{tt}^{\prime
}-\tilde{g}_{tt}\tilde{g}_{\varphi \varphi }^{\prime }}}\left( \sqrt{\tilde{g}_{\varphi \varphi
}^{\prime }}\pm \sqrt{\frac{\tilde{g}_{\varphi \varphi }}{\tilde{g}_{tt}}\tilde{g}_{tt}^{\prime }}%
\right).  \label{shift2}
\end{equation}%

In the unitary gauge we have $\phi(r) = \kappa^{-1}_{4}$, and under the conformal transformation the scalar field reads $\phi^{*}(r) = \Omega^{-1/2}\kappa^{-1}_{4}$. For this election of the gauge, the tilded components of the metric read
\begin{equation}
   \tilde{g}_{tt}= \phi^{*2}g_{tt} = \Omega^{-1}\kappa^{-2}_{4} \Omega \hat{g}_{tt} = \kappa^{-2}_{4} \hat{g}_{tt}, \label{Tildegtt}
\end{equation}
and
\begin{equation}
   \tilde{g}_{\varphi\varphi} =\phi^{*2}g_{\varphi\varphi} = \Omega^{-1}\kappa^{-2}_{4} \Omega \hat{g}_{\varphi\varphi} = \kappa^{-2}_{4} \hat{g}_{\varphi\varphi}. \label{Tildegphi}
\end{equation}

Then, if we substitute Eqs. \eqref{Tildegtt} and \eqref{Tildegphi}, as well as $\phi^{*}(r) = \Omega^{-1/2}\kappa^{-1}_{4}$ into Eq. \eqref{shift2}, we obtain an expression for the frequency shift with the explicit dependence of the spacetime parameters
\begin{eqnarray}
    &1+z^{CC}_{_{1,2}} = \sqrt{\frac{r_{e}}{\Omega(r)(r_{e}-3M)}}\left(1\pm \sqrt{\frac{M}{r_{e}-2M}}\right)\\
    &= \left(\frac{r^2}{r^2+l^2}\right)^{N}\sqrt{\frac{r_{e}}{(r_{e}-3M)}}\left(1\pm \sqrt{\frac{M}{r_{e}-2M}}\right)\notag,
\end{eqnarray}
where, as in the previous Section, the total frequency shift is composed by the gravitational frequency shift
\begin{equation}
\begin{split}
    1+z_{g}^{CC}&=\sqrt{\frac{r_e}{\Omega(r_e-3M)}},\label{zgravcc}
\end{split}
\end{equation}
and the kinematic frequency shift
\begin{equation}
\begin{split}
    z_{kin_{\pm }}^{CC}&=\pm\sqrt{\frac{Mr_e}{\Omega(r_e-3M)(r_e-2M)}},
\end{split} \label{zkincc}
\end{equation}
satisfying the same Eq. (\ref{ztotal}).

It is important to remark that for this gauge, several properties of the spacetime remain the same as in the Schwarzschild case, for instance, the light bending parameter, the effective potential, as well as the ISCO radius $r^{CC}_{ISCO} = 6M$.

On the other hand, due to the recession of the galaxy NGC 4258 with respect to us, it is necessary to take this motion into account, as it contributes to the frequency shift of photons emitted by masers. Thus, we model this recession motion with a special relativistic boost \cite{ApJLNucamendi}
\begin{equation}
    1+z_{boost} = \gamma(1+\beta),
\end{equation}
where $\gamma \equiv  (1-\beta^2)^{-1/2}$ and $\beta = v_p/c$, $v_p=z_p\,c$. Here, $v_p$ is the peculiar velocity of the galaxy with respect to a far away located observer, and the associated peculiar redshift is denoted by $z_p$.
Hence, the total frequency shift reads
\begin{equation}
    1+z_{tot_{1,2}}= ( 1+z_{_{1,2}})(1+z_{boost}),
\end{equation}
and the relativistic boost must be applied to consider a proper analysis of the redshift/blueshift for both conformally and non-conformally coupled particles.

\section{Observations of megamasers on the NGC 4258 system}
\begin{table*}[tbh]
\begin{center}
\begin{tabular}{ l  c  c  c  c  c }
\hline\hline Model & $M/D\,(\times 10^{6}M_{\odot}\text{Mpc}^{-1})$& $x_0$ (mas) & $z_p$ ($\times 10^{-3}$)  & $l/D\, (\text{ m }\text{Mpc}^{-1})$ & $\chi^2$ \\ &  &  & $v_p\,(\text{km}\,\text{s}^{-1})$ &  & \\ \hline\hline
NCC  ($N=1$) & $5.5881 \pm 0.0097$ & $-0.1809 \pm 0.0091$ & $1.572 \pm 0.001$ & $ (4.5049^{+0.073}_{-0.075})\times10^{11}$ & $1.1762$ \\ &  &  & $ 471.31\pm 0.58$ &  & \\ \hline
NCC ($N=2$) & $5.5882 \pm 0.0097$ & $-0.1809^{+0.0092}_{-0.0091}$ & $1.572 \pm 0.001$ & $  (3.1853^{+0.052}_{-0.053})\times10^{11}$ & $1.1762$ \\ &  &  & $ 471.30 \pm 0.58$ &  & \\ \hline
CC ($N=1$) & $5.3391 \pm 0.0021$ & $0.4957^{+0.0146}_{-0.0148}$ & $1.532 \pm 0.002$ & $ (9.9506^{+0.0630}_{-0.0650})\times10^{12}$ & $1.0977$ \\ &  &  & $ 462.33\pm 0.59$ &  & \\ \hline
CC ($N=2$) & $5.3390 \pm 0.0021$ & $0.4956^{+0.0147}_{-0.0149}$ & $1.532 \pm 0.002$ & $  (7.0353^{+0.0449}_{-0.0463})\times10^{12}$ & $1.0977$ \\ &  &  & $ 462.33\pm0.59$ &  & \\ \hline
Schwarzschild & $5.2990 \pm 0.0019$ & $-0.382^{+0.0049}_{-0.0050} $ & $1.615 \pm 0.001$ & - & $1.2077$ \\ &  &  & $484.31 \pm 0.35$ &  &\\
\hline\hline
\end{tabular}
\end{center}
\caption{Most probable values of the statistical fit for the BH parameters from the general relativistic model for nonsingular BHs in conformal gravity with conformal/non-conformal coupling in the particles' geodesics. Flat prior distributions have been set in appropriate intervals.}
\label{tab:Estimations}
\label{Results}
\end{table*}
\begin{figure*}[tbh]
    \centering
    \includegraphics[width=0.45\textwidth]{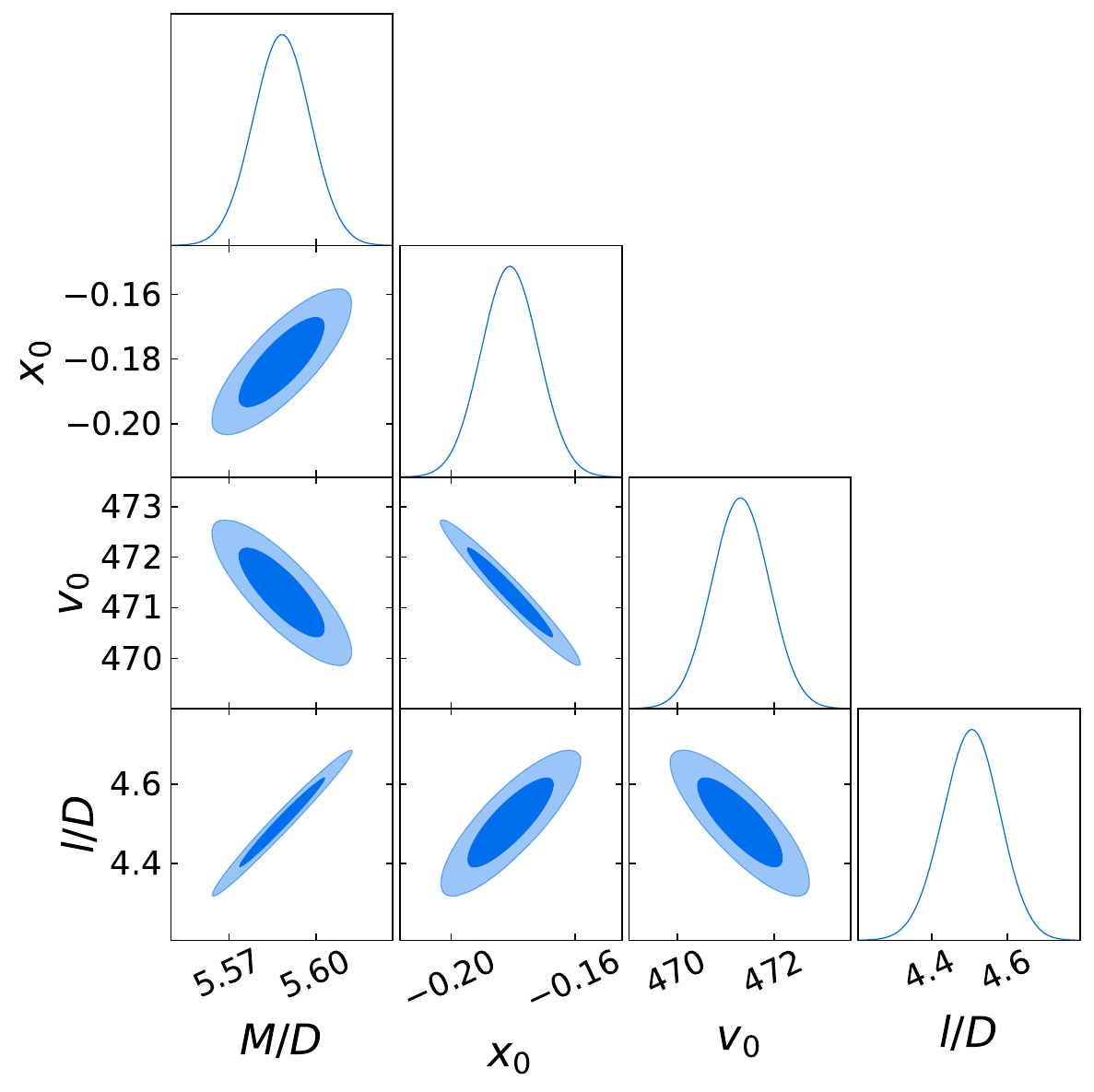}
    \includegraphics[width=0.45%
    \textwidth]{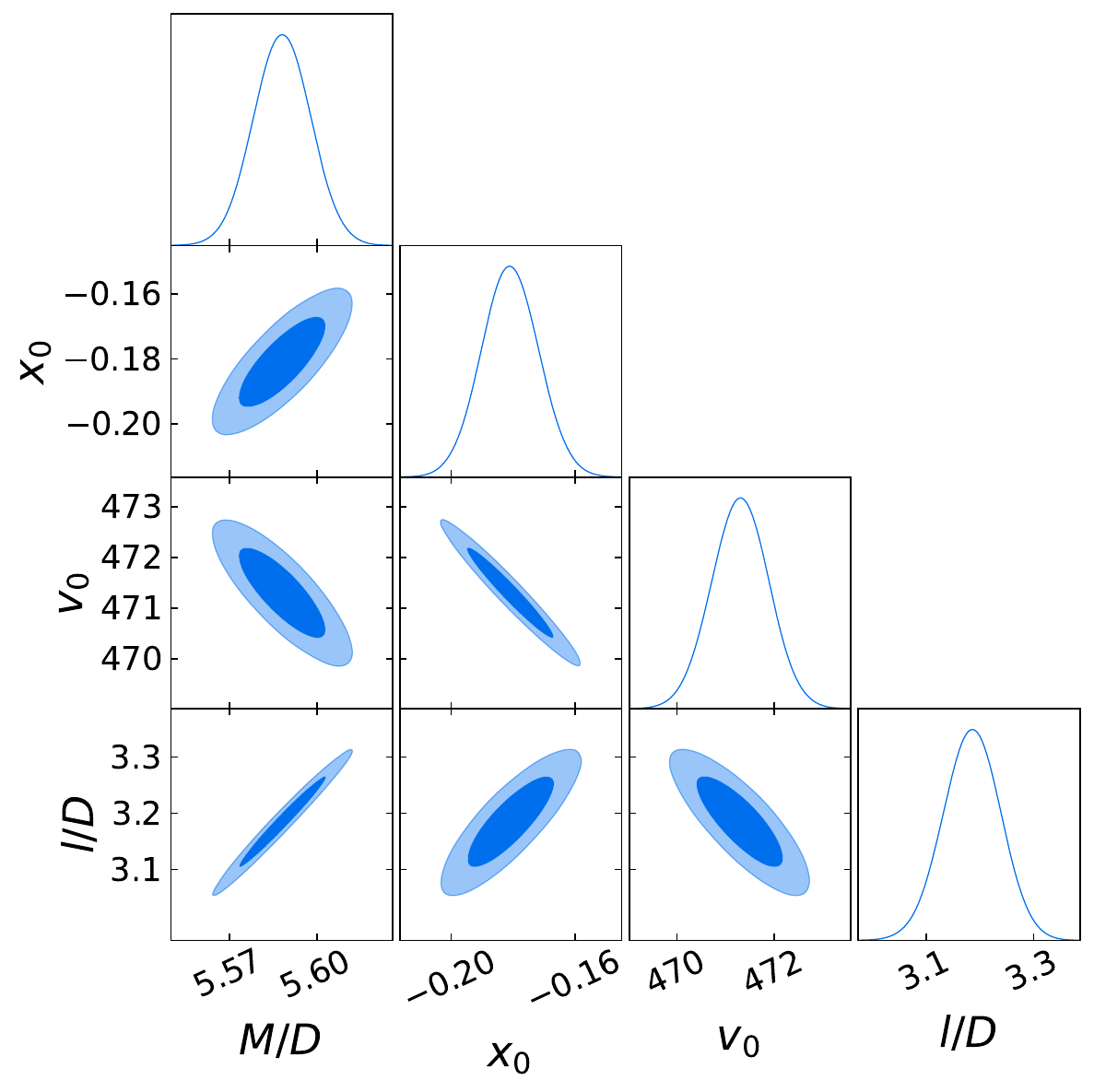}
    \includegraphics[width=0.45%
    \textwidth]{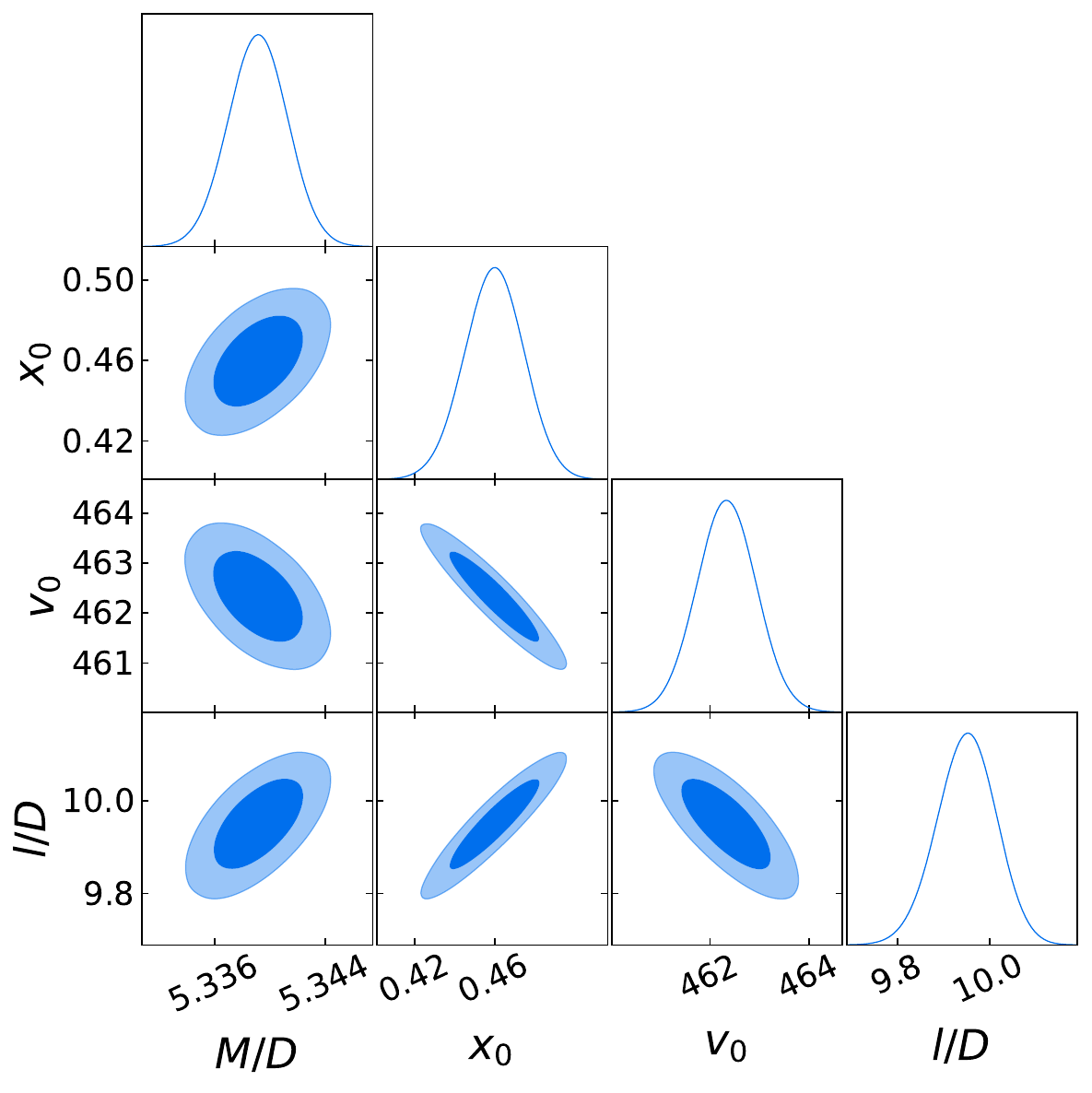}
    \includegraphics[width=0.45%
    \textwidth]{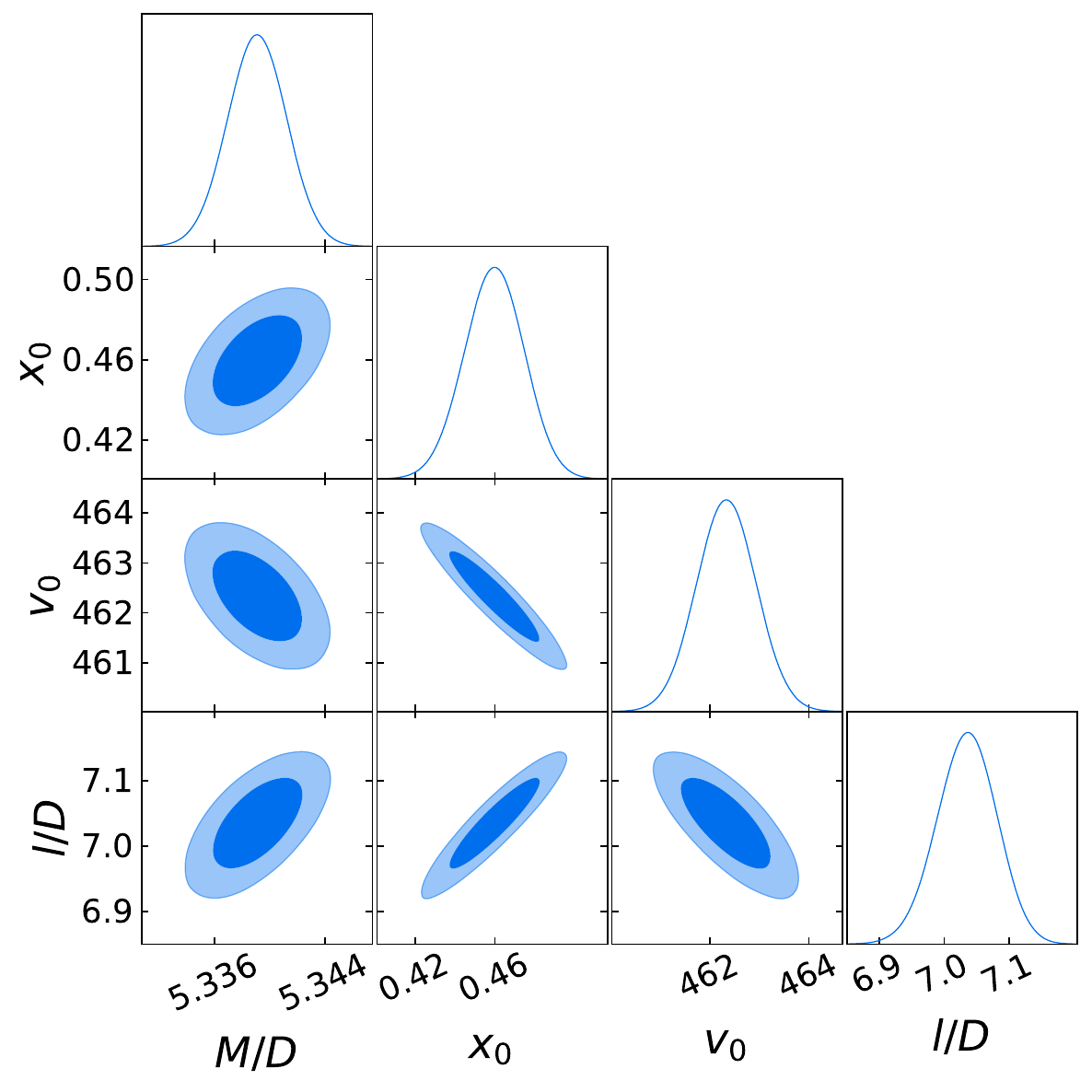}
    \caption{Posterior distribution of the conformal gravity fit  for NCC particles (upper left/right panel corresponds to $N=1$/$N=2$) and CC particles (lower left/right panel corresponds to $N=1$/$N=2$). Note that contour levels amount to $1\sigma$ and $2\sigma$ confidence regions.}
    \label{gaussian}
\end{figure*}

The analyzed astrophysical system consists of a subparsec, warped, rotating accretion disk containing several water masers in circular orbits viewed nearly edge-on from Earth within the NGC 4258 AGN hosting a SMBH. We make use of the 3 year 18 epoch VLBI observations of the megamaser system reported by Argon et al. \cite{Argon} that account for the redshifts/blueshifts of light particles emitted by the water masers at the points at which the velocity gain paths of the
probe particles are the longest, where $\varphi\approx\pm \pi/2$, allowing for a small scattering azimuthal angle around the midline. Measurements were taken with an adopted maser position $\alpha_{2000}=12^{\text{h}}18^{\text{m}}57.^{\text{s}}5046\pm0.^{\text{s}}0003$, $\delta_{2000}=47^{\circ}18'14.''303\pm0.''003$ \cite{NGC4258}. The NGC 4258 system was observed 18 times from 1997 March 6 through 2000 August 12, using the VLBA, the Effelsberg 100 m telescope (EFLS) of the Max-Planck-Institut für Radioastronomie, and the Very Large Array (VLA). Twelve ``medium-sensitivity'' epochs were carried out with the VLBA alone; while six ``high-sensitivity'' epochs were interleaved and involved the VLBA, enhanced by the phased VLA and EFLS. Calibration and synthesis imaging were performed using the Standard Astronomical Image Processing System (AIPS) package. The VLBA provides an angular resolution of 200 $\mu$as and a spectral resolution of 1 km s$^{-1}$ at the water maser frequency of 22 GHz \cite{Argon}.

\label{Observations}

\section{Estimation of nonsingular BH parameters from NGC 4258 data}

\label{nonsingular}

As it is shown in the previous Section, the megamaser system NGC 4258 meets the necessary conditions imposed in our formalism, namely, emitters in circular and nearly equatorial orbits. Therefore, by applying the formalism presented in Section \ref{GR model} to the spacetime \eqref{CG metric} we obtain the expression in Eq. \eqref{shift2} for the frequency shift of photons emitted by massive particles orbiting the BH \cite{MartínezV}. In this model, we consider that masers are scattered about the midline by an angle $\delta\varphi = 4.2^{\circ}$, and that the disk possesses an inclination of $82^{\circ}$ indicated by the polar angle $\theta_{0}$ \cite{ImprovedNGC4258,NGC4258}.

The parameters that undergo the statistical fit are the BH mass-to-distance ratio $M/D$, the conformal gravity length scale parameter-to-distance ratio $l/D$, the peculiar frequency shift $z_p$ and the horizontal offset $x_0$ of the center of the BH position $r_0(x_0,y_0)$. Furthermore, we see that the thin character of the accretion disk leads to an insensitive fit of the $y_0$ offset; a fact that allows us to fix $y_0=0.556$ mas \cite{ApJLNucamendi} following \cite{NGC4258} .

In addition, we minimize the $\chi^2$ by a Bayesian statistical fit making use of the Markov-Chain Monte Carlo (MCMC) algorithm, applying it to the rotation curves of the megamasers using the results from Section \ref{GR model}. Further, we obtain the errors $\sigma_{g}$ and $\sigma_{kin}$ associated with the gravitational and kinematic frequency shifts, respectively. For the non-conformally coupled particle these errors read
\begin{equation}
\medmath{
\begin{split}
    &\sigma_g^{NCC}=-\frac{1}{2 r_e}\left(\frac{1}{\Omega(1-3\tilde{M})^3(1+A)}\right)^{1/2}\\
    &\times\left(\frac{2A}{\left(1+\tilde{l}^2\right)}(1-3\tilde{M})+(1+A)\left[2A(1-3\tilde{M})+3\tilde{M}\right]\right),
\end{split}}
\end{equation}
and
\begin{equation}
\medmath{
\begin{split}
      &\sigma_{kin}^{NCC}=-\frac{\cos{(\delta\varphi)}\sin{\theta_0}}{2 r_e}\left(\frac{1}{\Omega(1-3\tilde{M})^3(A+B)}\right)^{1/2}\times\\
      &\left(\left[\frac{2A}{\left(1+\tilde{l}^2\right)}+\frac{B}{(1-2\tilde{M})}\right](1-3\tilde{M})\right.\\
      &\left.+(B+A)\left[2A(1+3\tilde{M})+3\tilde{M}\right]\right), 
\end{split}}
\end{equation}
where $A=-\frac{2N\tilde{l}^2}{1+\tilde{l}^2}$, $B= \frac{\tilde{M}}{1-2\tilde{M}}$,  $\tilde{M}=\alpha\frac{M}{r_e}$ and $\tilde{l}=\lambda\frac{l}{r_e}$, whereas for the conformally coupled probe particle we have the following expressions
\begin{eqnarray}
    \sigma_{g}^{CC} = \frac{2N\tilde{l}^2}{r_{e}^2(1+\tilde{l}^2)}(1+z_g)-\frac{3}{2}\frac{\tilde{M}}{r_{e}}\Omega(1+z_{g})^3,
\end{eqnarray}
and
\begin{equation}
    \begin{split}
    \sigma_{kin}^{CC} &= \frac{\cos{(\delta\varphi)}\sin{\theta_0}}{2}\left[\Omega z^3_{kin}\left(\frac{6\tilde{M}^2-1}{\tilde{M}r_{e}}\right)\right.\\
    &+z_{kin}\left.\frac{4N\tilde{l}^2}{r_{e}(1+\tilde{l}^2)}\right].
\end{split}
\end{equation}

Since the metric \eqref{CG metric} is expressed in geometrized units, we introduce the factor $\alpha=\frac{GM_\odot\lambda}{c^2}$ to convert mass and distance from geometrized units to SI units, where $\lambda^{-1} = (4.848\times10^{-9})(3.086\times10^{22})$ is the conversion factor (from mas to radians and Mpc to meters) for the radius of the emitter $r_e$.

Following previous works \cite{ApJLNucamendi,Villaraos,NGC4258,AGJ}, and from the information given above, we can construct an expression for $\chi^2_{_{\pm}}$ of the general relativistic model as follows
\begin{eqnarray}
&&\chi^2_{_{\pm}}= \\
&&\medmath{\displaystyle\sum_{i=1}^{N} \frac{\left[z_{obs,i}-\left((1+z_{g}+ z_{kin_{\pm}}\cos{(\delta\varphi)}\sin{\theta_0})(z_{boost}+1)-1\right)\right]^2}{ \sigma_{z_{i}}^2 + \sin^2\theta_0(z_{kin_{\pm}} (1+z_{boost}))^2 (\delta\varphi)^2\,\sin^2{(\delta\varphi)}}} \notag
    \label{Chi Squared},
\end{eqnarray}
where the summation is performed over the measured frequency shift of the $i$-th maser feature, and
\begin{equation}
\begin{split}
    \sigma_{z_{i}} &= (\sigma_g + \sigma_{kin_{\pm}})(z_{boost}+1)\delta r_e,
\end{split}
\end{equation}
\begin{equation}
    \delta r_e = D \delta\Theta,
\end{equation}
\begin{equation}
     \delta\Theta = \left[ \left( \frac{(x-x_0)\delta x}{\Theta} \right)^2 + \left( \frac{(y-y_0)\delta y}{\Theta} \right)^2 \right]^{1/2}, \label{Rerror}
\end{equation}
\begin{equation}
    \Theta = \sqrt{(x-x_0)^2 + (y-y_0)^2},
\end{equation}
where $\Theta$ is the angle between a given maser and the BH.

By performing the statistical fit, we obtain the values for the BH parameters registered in Table \ref{tab:Estimations} for the fixed values of $N=1,2$. This amounts to choosing two different conformal factors $\Omega_{_{N=1}}$ and $\Omega_{_{N=2}}$ in Eq. \eqref{CG metric}, and once we fixed the value of parameter $N$, we focus on estimating the ratio $l/D$, which is the remaining parameter of interest in the conformal gravity theory. From Table \ref{tab:Estimations}, we see that the estimation of the parameters $M/D$, and $z_p$ render very similar results to those obtained in the case of the Schwarzschild model, which, as it can be readily verified in Eq. \eqref{CG metric}, amounts to $N=0$. Besides, although the estimation for the BH position $x_0$ within the conformal gravity theory deviates from the one obtained in the Schwarzschild model, it is still physically viable given the distribution of the masers on the sky. In addition, we observe from Fig. \ref{gaussian} that the MCMC method applied to the data set using our general relativistic model leads to a posterior Gaussian distribution for all the parameters. Further, by comparing the cases $N=1$ and $N=2$, we see that the posterior results are very close to each other, with a reduced $\chi^2=1.1762$ for NCC particles, and $\chi^2 = 1.0977$ for CC particles, in both cases (see Table \ref{tab:Estimations}). Besides,  the estimations of the ratio $l/D$ in non-conformal coupling for $N=1,2$ are relatively close and of the same order when compared to each other. Similarly, conformally coupled geodesics render estimations of the parameter $l/D$ of the same order among $N=1,2$, but are approximately ten times greater than NCC particles. Additionally, the remaining BH parameters stay similar between conformal and non-conformal coupling scenarios. Furthermore, finding a most probable value for $l$ with a Gaussian distribution provides a strong evidence for a nonvanishing length scale parameter. Hence, these results assert that the conformal gravity theory is suitable for describing a nonsingular BH at the core of the NGC 4258 AGN according to the general relativistic model presented in Section \ref{GR model}.

\section{Discussion and final remarks}

In this work, we performed a Bayesian statistical fit using the MCMC method and observational data from the archetypal megamaser system NGC 4258 (the megamaser system with the largest data set so far) to estimate the spacetime parameters of a nonsingular BH within the framework of conformal gravity. To achieve this, we made use of a general relativistic method developed in previous works, using the metric that describes a static spherically symmetric BH in conformal gravity. This BH possesses three parameters that characterize the background spacetime, a length scale parameter $l$, an integer parameter $N$ which denotes a family of conformal transformations $\Omega_{_{N}}$, and the mass of the BH $M$. We used the MCMC method with a large number of chains ($10^{5}$) and computational resources from the Laboratorio Nacional de Supercómputo del Sureste de México (LNS) in order to obtain a precise value for the estimation of the ratio $l/M$ which provides additional information with respect to the one obtained in previous studies. We can appreciate the confidence regions for the statistical fit from Fig. \ref{gaussian}, and using prior flat distributions in appropriate intervals, the posterior distribution renders a Gaussian distribution for all the parameters.

\begin{figure}[tbh]
    \centering
    \includegraphics[width=0.5\textwidth]{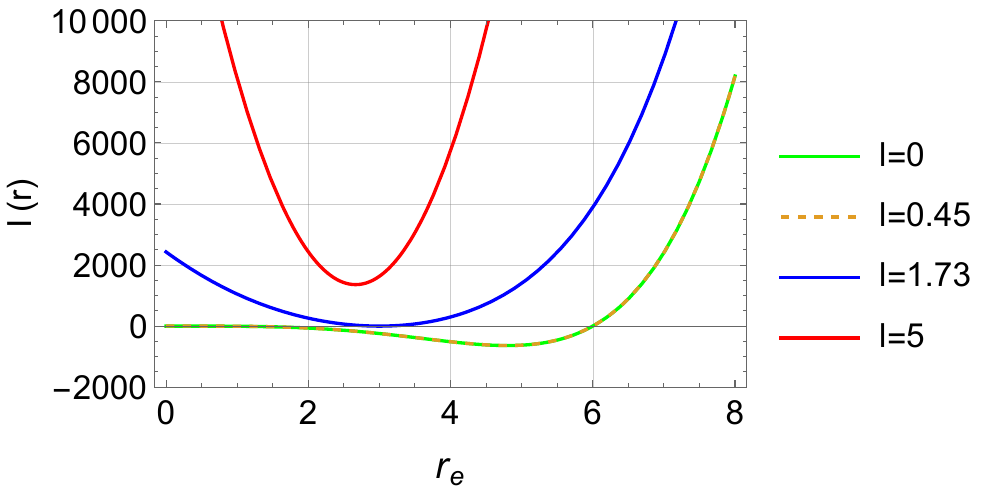}
    \includegraphics[width=0.5%
    \textwidth]{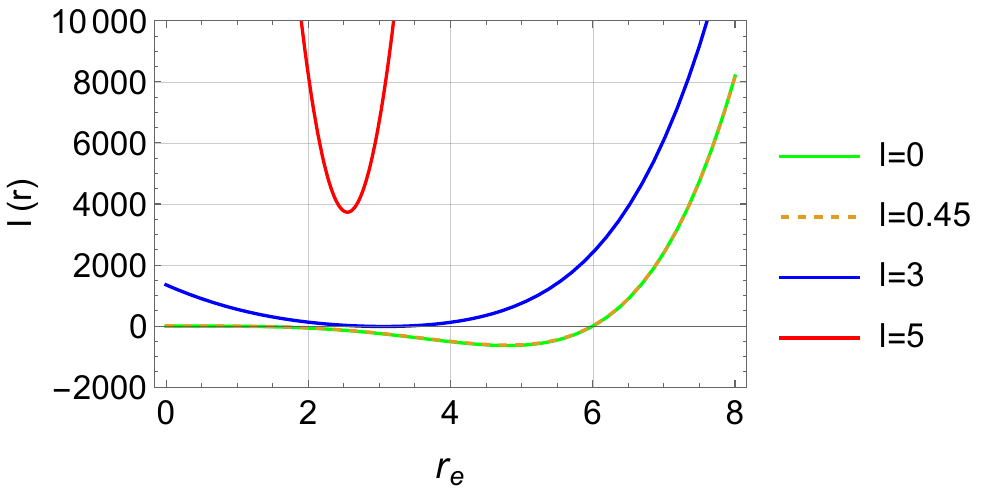}
    \caption{Plot of the ISCO Eq. \eqref{rISCOeq} for $N=1$/$N=2$ (top/bottom), in units where $M=1$. Here it is shown that for small values of $l$, the $r_{ISCO}$ approaches the Schwarzschild $r_{ISCO}$  ($r=6$), and for larger values of $l$, the $ISCO$ curves does not intersect the horizontal axis.}
    \label{ISCOeq}
\end{figure}

It is worth mentioning that in our formalism, the mass and the length parameter are coupled to the distance $D$ between the observer and the BH. Thus, we estimate the ratios $M/D$ and $l/D$, and divide them to obtain the $l/M$ ratio, so that we can compare it with previous constraints \cite{TestCG1,TestCG2,TestCG3}.

With these estimations we have obtained accurate predictions of the values for the ratio $l/D$. Furthermore, we compute the values of the ratio $l/M$ (in geometrized units), and the parameter $l$, for $N=1,2$ with their associated error (see Table \ref{tab:Conversions}), by taking into account the results from Table \ref{tab:Estimations}, as well as the distance for NGC 4258 obtained in \cite{ImprovedNGC4258}.

\begin{table*}[tbh]
\begin{center}
\begin{tabular}{ l  c  c  c  c }
\hline\hline Coupling & $(l/M)_{N=1}$ & $(l/M)_{N=2}$ & $l_{N=1}$ (m) & $l_{N=2}$ (m) \\ \hline\hline
NCC & $54.5513 \pm 0.9010$ & $38.5719 \pm 0.6332$ & $ (3.4129 \pm 0.0056)\times10^{12}$ & $(2.4131 \pm 0.0039)\times10^{12}$ \\ \hline
CC & $1261.15 \pm 8.25$ & $891.65 \pm 5.70$ & $ (7.53857 \pm 0.0049)\times10^{13}$ & $(5.32994 \pm 0.0034)\times10^{13}$ \\ \hline\hline
\end{tabular}
\end{center}
\caption{Most probable values for the ratio $l/M$ and the parameter $l$ with their associated errors. These values have been obtained by considering the estimations of $M/D$ and $l/D$ from Table \ref{tab:Estimations} (to estimate $l/M$), and the value of $D=7.576\pm 0.082$ \cite{DistanceNGC4258} to estimate $l$. The corresponding errors have been properly propagated.}
\label{tab:Conversions}
\label{Results}
\end{table*}

We remark that we have obtained new results in contrast to previous constraints, and from a reliable model, using BH rotation curves. Moreover, in comparison to previous works, where only upper bounds were found, here we obtain the most probable value for the parameter $l$ in a Gaussian distribution, which indicates that nonvanishing values for the length scale parameter can suitably model the observations from the NGC 4258 AGN. 

\begin{figure}[tbh]
    \centering
    \includegraphics[width=0.4\textwidth]{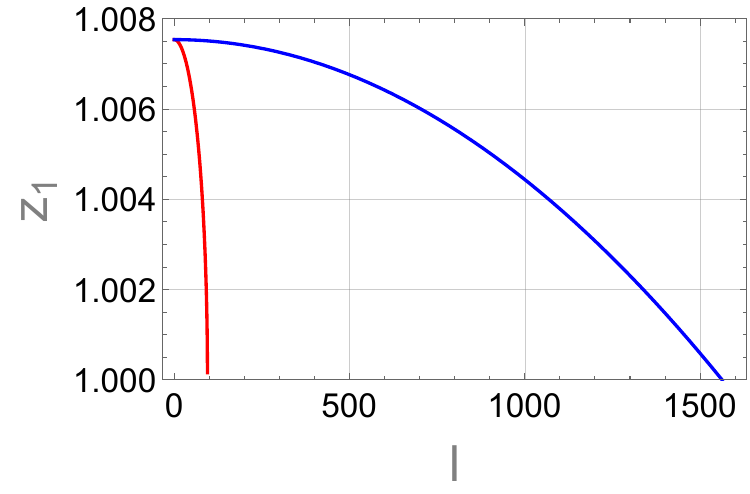}
    \includegraphics[width=0.4\textwidth]{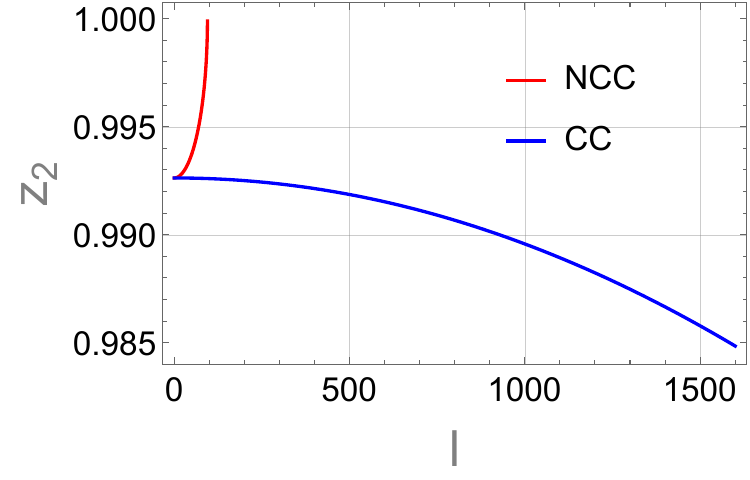}
    
\caption{Plots of the total redshift/blueshift for conformally and non-conformally coupled particles as a function of $l$ for fixed values of $r_{e} = 1$ pc, $M=1$ and we display only the case $N=1$ for ilustrative purposes. It can be observed that the NCC frequency shift decays faster, and therefore, it takes greater values of $l$ for CC to reach the same values.}
\label{Decay}
\end{figure}

As illustrated in Table \ref{tab:Estimations}, the value of the length scale parameter for the CC particle case is greater than the one obtained in the NCC case, as expected from analyzing the expressions \eqref{Shift1} and \eqref{shift2}. We can see that both expressions are rescaled by a factor $\Omega^{-1/2}$, while for Eq. \eqref{Shift1} there is an additional effect of the conformal parameters in the gravitational and kinematic frequency shift due to the term $-2Nl^2/(r^2+l^2)$, which causes the expression for $z_{_{1,2}}$ to decay faster as a function of $l$, as shown in Fig. \ref{Decay}.

In addition, our results on the estimations of the length scale parameter for non-conformally coupled massive test particles lead to new interesting physical properties resulting from an analysis of the solutions to the $r_{ISCO}$ equation $I(r)$, defined as follows \cite{MartínezV}
\begin{equation}
\begin{split}
&I(r)=r^{5}-6Mr^{4}-2l^{2}(N-1)r^{3}+  \\
&+\frac{4l^{2}}{M}\left[Nl^{2}\left( 2N-1\right) +3M^2(N-1)\right] r^{2}-  \\
&l^{4}\left( 2N-1\right) \left[ (20N+1)r-6M(4N+1)\right]=0,\end{split}
 \label{rISCOeq}
\end{equation}
which has no physical solutions for any of the most probable values $(l/M)_{_{N=1}}$ and $(l/M)_{_{N=2}}$ from Table \ref{tab:Conversions}. The numerical solutions to Eq. \eqref{rISCOeq} for the aforementioned values of $N$ and $l$ render either negative or complex conjugated roots.
\begin{figure}[tbh]
    \centering
    \includegraphics[width=0.482\textwidth]{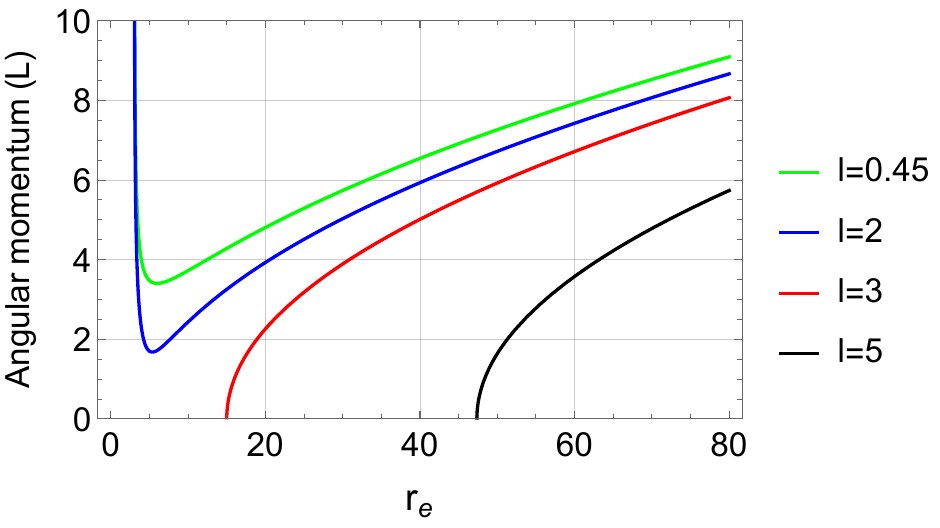}
    \includegraphics[width=0.48\textwidth]{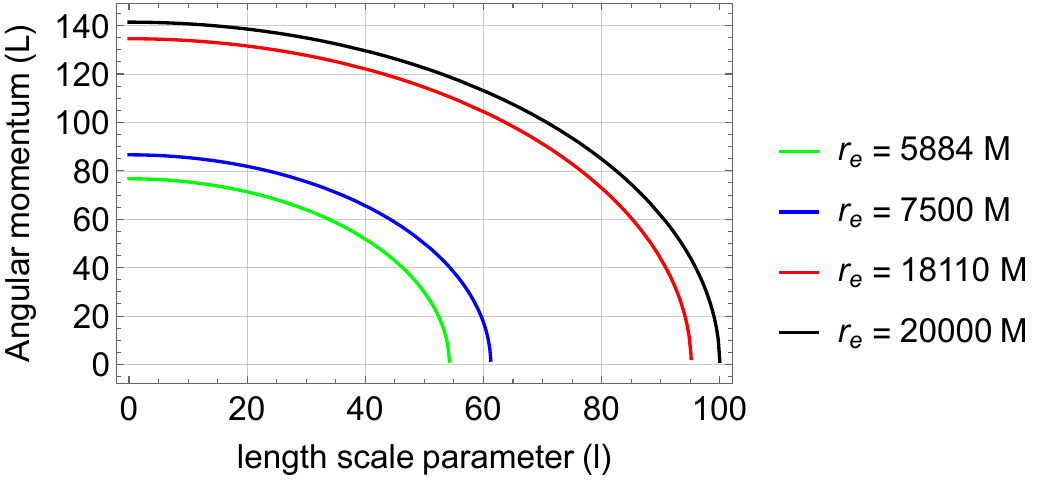}   
\caption{Plots of the angular momentum $L$ expression in Eq. \eqref{AngularMomentum} for $N=1$ in units where $M=1$, as a function of $l$ (top) and as a function the radius of the emitter $r_e$ (bottom).}
\label{AngMomPlot}
\end{figure}
Furthermore, an analysis of $V_{eff}$ as a function of $l$ in the vicinity of $r_{e}=3M$ reveals that at the critical ratio $(l/M)_{c} = \frac{3}{\sqrt{2N-1}}$, the effective potential stops having a minimum, and consequently, for $l/M\geq (l/M)_{c}$ there is no $r_{ISCO}$. This behavior is illustrated in Fig. \ref{ISCOeq}, where we see that as the parameter $l$ increases, the $ISCO$ curves depart from the horizontal axis, and the blue curves correspond to the critical ratio ($(l/M)_{c}=3$ for $N=1$ and $(l/M)_{c}=\sqrt{3}\approx1.73$ for $N=2$).

Additionally, for a better understanding of the limits of the circular orbital motion, we study the expression for the angular momentum of the emitters
\begin{equation}
    L = \pm\sqrt{\frac{\Omega r^2\left[M(r^2+l^2)-2Nl^2(r-2M)\right]}{(r-3M)(r^2+l^2)}}. \label{AngularMomentum}
\end{equation}

\begin{figure*}[tbh]
\centering
\includegraphics[width=0.49%
\textwidth]{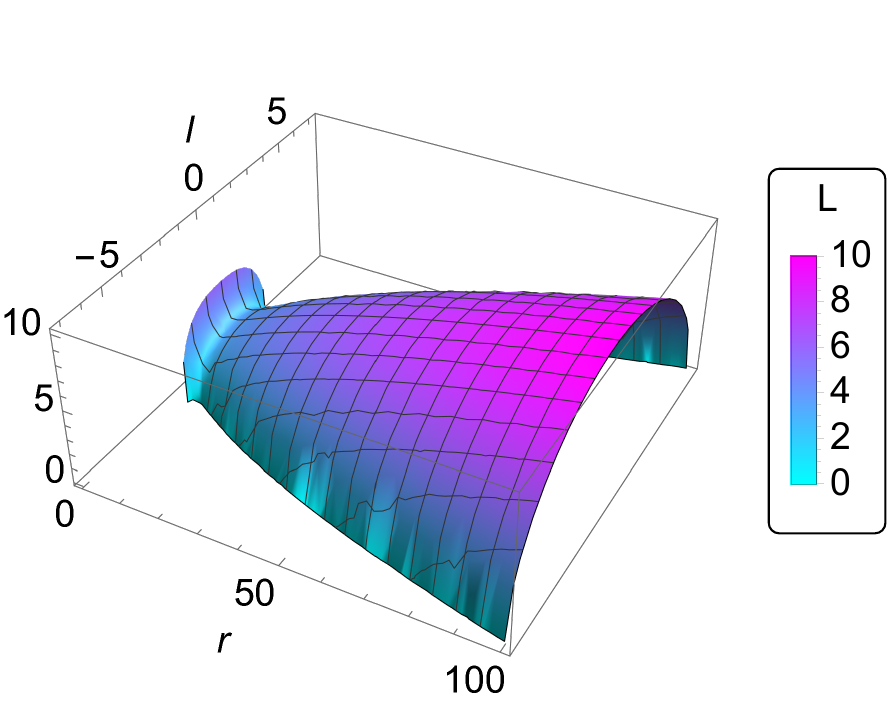}
\includegraphics[width=0.49%
\textwidth]{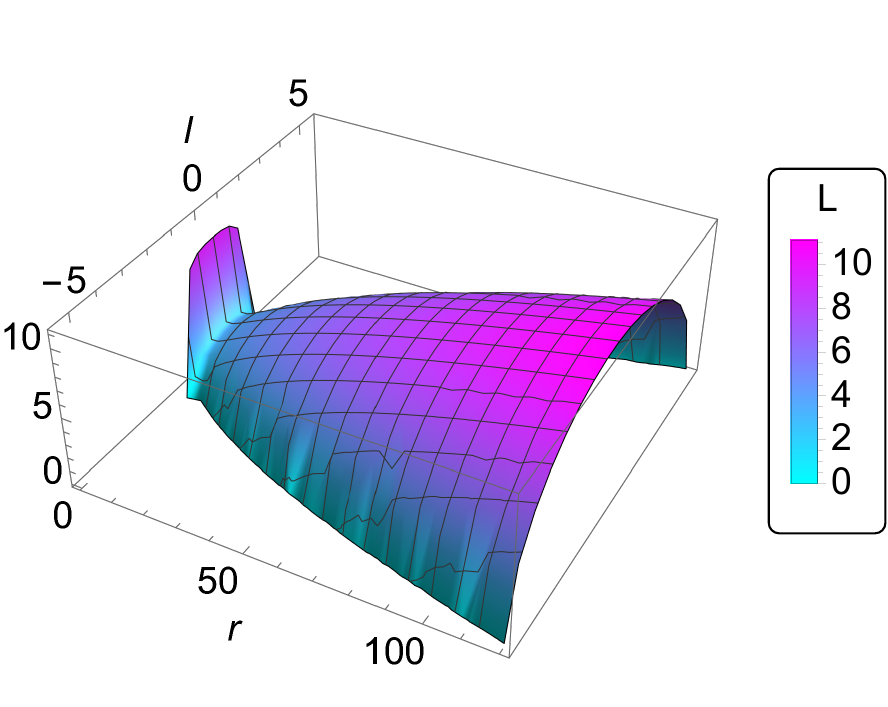}
\caption{Plots of the angular momentum $L$ expression from Eq. \eqref{AngularMomentum} for $N=1$/$N=2$ (left panel/right panel), in units where $M=1$, as a function of $l$ and the radius of the emitter $r_e$.}
\label{AngMomPlot2}
\end{figure*}

The first remarkable feature of this expression is the decreasing character of the angular momentum as a function of the conformal parameters. In particular, as the length scale parameter increases, $L$ decreases as illustrated in Fig. \ref{AngMomPlot} (top panel). Secondly, we find that $L$ is not defined for the region $r\in(r_{-},r_{+})$, where
\begin{equation}
    r_{\pm} = \frac{Nl^2\pm l\sqrt{N^2l^2-M^2(1+4N)}}{M}, \label{roots}
\end{equation}
and therefore, $r_{+}$ represents a lower bound for the particles' orbits.

From this expression, we observe that the discriminant is positive if $l/M \geq \frac{\sqrt{1+4N}}{N}$. This means that values of $l/M$ smaller than this lower bound lead to a negative discriminant. In such a case, the roots $r_{\pm}$ are complex-conjugated, and therefore the expression for the angular momentum \eqref{AngularMomentum} is well-defined for $r\in(3M,\infty)$, as illustrated in Fig. \ref{AngMomPlot} for the blue and green curves.  On the other hand, the red and black curves correspond to the case of $l/M$ larger than the aforementioned lower bound, which, for $N=1$ reads $l/M\geq \sqrt{5}$. The red and black curves intersect the horizontal axis at the points $r^{(red)}_{+} = 15 M$ and $r^{(black)}_{+} = 47.36 M$, respectively.

In this regard, taking into account our estimations of $l/M$, Eq. \eqref{AngularMomentum} yields a minimum radius for the orbits of the particles, which, for the case of $N=1$ is $r_{+}=5884.04 M$, while for $N=2$, we obtain $r_{+} = 5888.31 M$. If we compare these values with the radius of the closest maser to the BH, $r_{e} \approx 0.001 $ mas $ = 1.811\times10^{4} M$, we observe that the minimum value of $r$ is relatively close to the radius of the emitter for the NGC 4258 system. Hence, even though the ISCO equation has no physically meaningful roots for our estimations of the $l/M$ ratio, the orbital radius is still bounded by $r_+$.

A more complete description of this behavior is provided in Fig. \ref{AngMomPlot2}, where we observe the combined effects of increasing $l$ and $r_{e}$ on the angular momentum $L$. The regions where the angular momentum vanishes correspond to large values of $l$, more precisely, for $l>\sqrt{5}$ for $N=1$ (left panel) and $l>1.5$ for $N=2$ (right panel). We also note that as $l$ increases, the point where $L$ vanishes (which amounts to $r_{+}$) also increases, as expected from Eq. \eqref{roots}. 

Further work includes considering a more general spacetime with a cosmological constant, removing the singularity at $r=0$ by applying a conformal transformation to the Schwarzschild-de Sitter spacetime, which allows the use of data from active galaxies within the Hubble flow. This extension of the model permits us to compare estimations for the $l/D$ parameter using several galaxies, and allows one to use the relation between the distance $D$ and the Hubble constant $H_{0}$ to decouple $D$ from the mass and the length scale parameter in such a way that we can estimate separately the $M$ and $l$ parameters.

\label{Final remarks}

\section*{Appendix: Conformally invariant, weakly non-local, UV finite theory}
\label{appendix}
Here we discuss the main ideas and subtleties from the conformally invariant, weakly non-local and UV finite theory presented in \cite{Modesto16}.
We originally consider the following action
\begin{equation}
    \mathcal{L}_{g}=-2\kappa^{-2}_{D}\sqrt{g}\left(R + R_{\mu\nu\rho\sigma}\gamma(\square)^{\mu\nu\rho\sigma}_{\alpha\beta\tau\delta}R^{\alpha\beta\tau\delta}+V\right), \label{generalAction}
\end{equation}
where $\kappa_{D}$ is the gravitational coupling constant in $D$ dimensions, the Riemann tensor is defined by 
\begin{equation}
    R^{\mu}_{\,\,\,\nu\rho\sigma} = \partial_{\rho}\Gamma^{\mu}_{\nu\sigma}-\partial_{\sigma}\Gamma^{\mu}_{\nu\rho} +\Gamma^{\mu}_{\tau\rho}\Gamma^{\tau}_{\nu\sigma} - \Gamma^{\mu}_{\tau\sigma}\Gamma^{\tau}_{\nu\rho},
\end{equation}
and the $\gamma(\square)$ is a weakly non-local function of the d'Alembertian operator $\square = g^{\mu\nu} \nabla_{\mu}\nabla_{\nu}$, defined as follows
\begin{equation}
    \gamma(\square)^{\mu\nu\rho\sigma}_{\alpha\beta\tau\delta} = g^{\mu\rho}g^{\nu\sigma}g_{\alpha\tau}g_{\beta\delta}\gamma_{0}(\square)+g^{\mu\rho}g_{\alpha\tau}\delta^{\nu}_{\beta}\delta^{\sigma}_{\delta}\gamma_{2}(\square),
\end{equation}
where the form factors are given by
\begin{equation}
    \gamma_{0}(\square) = -\frac{(D-2)(e^{H_{0}}-1)+D(e^{H_{2}}-1)}{4(D-1)\square},
\end{equation}
\begin{equation}
    \gamma_{2}(\square) = \frac{e^{H_{2}}-1}{\square},
\end{equation}
and $D$ is the dimension of the spacetime. Moreover, $e^{H_{0}}$ and $ e^{H_{2}}$ are dimensionless functions of the d'Alembertian operator, introduced to ensure UV finiteness and ghost freedom. Additionally, the potential $V$, constructed to be at least cubic in the curvature, reads
\begin{equation}
\begin{split}
        V=&\sum_{j=3}^{\mathcal{N}+2}\sum_{k=3}^{j}\sum_{i} c_{k,i}^{(j)} \left(\nabla^{2(j-k)}R^{k}\right)_{i}\\
        +&\sum_{j=\mathcal{N}+3}^{\tau+\mathcal{N}+1}\sum_{k=3}^{j}\sum_{i} d_{k,i}^{(j)} \left(\nabla^{2(j-k)}R^{k}\right)_{i}\\
        +&\sum_{k=3}^{\tau+\mathcal{N}+2}\sum_{i} s_{k,i} \left(\nabla^{2(\tau+\mathcal{N}+2-k)}R^{k}\right)_{i},
\end{split}
\end{equation}
where the coefficients $c_{k,i}^{(j)}, d_{k,i}^{(j)}$ and $s_{k,i}$ are coupling constants, and $\mathcal{N}$ is related to the spacetime dimension by  $D_{\text{even}}=2\mathcal{N}+4$ for even dimensions or $D_{\text{odd}}=2\mathcal{N}+3$ for odd dimensions. 

Then, in order to make the theory conformally invariant, we redefine the \textit{physical metric} $g_{\mu\nu}$ in terms of a scalar field $\phi$, and a \textit{background metric} $\hat{g}_{\mu\nu}$ as follows
\begin{equation}
    g_{\mu\nu} = (\kappa_{D}\phi)^{\frac{4}{D-2}} \hat{g}_{\mu\nu},\label{physicalmetric}
\end{equation}
where we take $\hat{g}_{\mu\nu}$ to be the Schwarzschild metric. Thus, it is clear that Eq. \eqref{physicalmetric} is invariant under the conformal transformations
\begin{equation}
    \hat{g}_{\mu\nu}^{*}=\Omega^2\hat{g}_{\mu\nu},\,\,\,\,\,\,\,\,\,\,\,\,\,\,\,\,\phi^{*}=\Omega^{\frac{2-D}{2}}\phi,
\end{equation}
and by substituting Eq. \eqref{physicalmetric} into Eq. \eqref{generalAction}, we obtain the following Lagrangian density
    \begin{equation}
    \begin{split}
         &\mathcal{L}_{g} = -2\sqrt{\hat{g}}\left[\phi^2R(\hat{g})+\frac{4(D-1)}{D-2}\hat{g}^{\mu\nu}\partial_{\mu}\phi\partial_{\nu}\phi\right]\\
         &-\frac{2}{k_{D}^2}\sqrt{g}\left[R(g)\gamma_{0}(\square)R(g)+R_{\mu\nu}(g)\gamma_{2}(\square)^{\mu\nu}_{\rho\sigma}R^{\rho\sigma}(g)\right.\\
         &\left.\left.+V(g)\right]\right|_{\phi\hat{g}},
    \end{split}   
\end{equation}
which is the conformally invariant version of the theory described by the lagrangian density in Eq. \eqref{generalAction}.

It is worth mentioning that the scalar compensator field $\phi$ does not account for an external field or matter, it is a gauge degree of freedom from the conformal symmetry. Therefore, this degree of freedom allows us to fix the gauge such that $\phi^{*\frac{4}{D-2}} \propto \Omega^2$.

Furthermore, if we investigate the dynamics derived from the action \eqref{generalAction} with respect to the physical metric $g_{\mu\nu}$, the compensator field $\phi$ does not give rise to any additional force, as it is coupled to the background metric $\hat{g}_{\mu\nu}$. On the other hand, if we compute the dynamics taking into account the metric $\hat{g}_{\mu\nu}$, additional forces appear. This is more clearly visualized when we relate the Christoffel symbols from the two frames as follows
\begin{equation}
    \Gamma(g)^{\mu\nu} =  \Gamma(\hat{g})^{\mu\nu} + (\text{terms involving }\partial \ln\phi),
\end{equation}
where the additional forces appear only in the non-physical frame of the background metric $\hat{g}_{\mu\nu}$, which carries no dynamical significance in our formalism.

\section*{Acknowledgements}
All authors are grateful to D. Villaraos, A. González-Juárez and M. Momennia for fruitful discussions and to FORDECYT-PRONACES-CONACYT for support under grant No. CF-MG-2558591; A.H.-A. thanks SNII
and was supported by a grant VIEP-BUAP. D.A.M.-V. acknowledges financial
assistance from CONAHCyT through the grant No. C-1325519.
The authors also thankfully acknowledge the computer resources, technical expertise, and support provided by the Laboratorio Nacional de Supercómputo del Sureste de México, CONAHCyT member of the network of national laboratories.\\


%

\printbibliography

%
%

\end{document}